\documentclass[letterpaper]{article}
\usepackage{aaai}
\usepackage{times}
\usepackage{helvet}
\usepackage{courier}

\usepackage{graphicx}

\title{The Call of the Crowd: Event Participation in Location-based Social Services}

\author{
Petko Georgiev\\
Computer Laboratory \\
University of Cambridge, UK\\
petko.georgiev@cl.cam.ac.uk\\
\And 
Anastasios Noulas \\
Computer Laboratory \\
University of Cambridge, UK\\
anastasios.noulas@cl.cam.ac.uk\\
\And
Cecilia Mascolo\\
Computer Laboratory \\
University of Cambridge, UK\\
cecilia.mascolo@cl.cam.ac.uk\\
}

\begin{document}
\maketitle
\begin{abstract}
\begin{quote}
Understanding the social and behavioral forces behind event participation is not only interesting from the viewpoint of social science, but also has important applications in the design of personalized event recommender systems.
This paper takes advantage of data from a widely used location-based social network, Foursquare, to analyze event patterns in three metropolitan cities. We put forward several hypotheses on the motivating factors of user participation and confirm that social aspects play a major role in determining the likelihood of a user to participate in an event. While an \emph{explicit} social filtering signal accounting for whether friends are attending dominates the factors, the popularity of an event proves to also be a strong attractor. Further, we capture an \emph{implicit} social signal by performing random walks in a high dimensional graph that encodes the place type preferences of friends and that proves especially suited to identify relevant \emph{niche} events for users. Our findings on the extent to which the various temporal, spatial and social aspects underlie users' event preferences lead us to further hypothesize that a combination of factors better models users' event interests. We verify this through a supervised learning framework. We show that for one in three users in London and one in five users in New York and Chicago it identifies the exact event the user would attend among the pool of suggestions.
\end{quote}
\end{abstract}

\section{Introduction}
Organized events such as festivals, concerts and sports games are important social phenomena offering individuals a source of recreation and opportunities to socialize. Understanding the collective dynamics of user participation in such events can provide critical insights that help in venue resource planning \cite{DBLP:conf/ht/LiangCCK13}, personalized event recommendation \cite{Minkov:2010:CFE:1871437.1871542} and targeted advertising that increase customer satisfaction and trust in online services. With the rise in popularity of location-based services such as Foursquare, we now have the tools to analyze and model social event participation at scale. The data from millions of users broadcasting their locations provides an unprecedented opportunity to accurately model the socio-spatial dynamics of events that motivate people to share their location and visit certain venues.

The present work studies the social and behavioral underpinnings of event participation as represented by location-based social networks. The main research question we address in this work is: \textbf{\emph{what is the extent to which geospatial, temporal, and social factors influence users' preferences towards events?}} To answer this question, we formulate a predictive modeling task where we try to match a user's mobility profile against the collective past check-in activity of potential event attendees. The design of this prediction task allows us to empirically measure homophily effects on users' event choices as reflected by location-based social media.

Besides its societal importance, solving the above mentioned challenge finds crucial applications in the domain of personalized event recommendations. First, the insights on the relationship between social media usage and event interests can be used to augment the credibility of recommendations by accompanying event suggestions with evidence elicited from publicly shared data. Further, under the assumption that some information about potential or on-going attendance is available, our framework can be directly used as a content-based event recommender system for users of location-based services. Such an assumption is being increasingly supported by the rapid growth of event-based social services provided by Facebook, Meetup, Plancast, DoubanEvent, and Eventbrite. These online networks offer a platform for users to not only organize and establish social events, but also to express their intention to join by signing up in advance. While event-based networks proliferate, Foursquare as a location-based service provides a unique chance to investigate event participation from multiple angles (temporal, spatial, geo-social) which are not simultaneously available elsewhere.

\begin{figure*}[!ht]  
        \begin{minipage}[b]{0.32\textwidth}
                \centering
                \includegraphics[width=\textwidth]{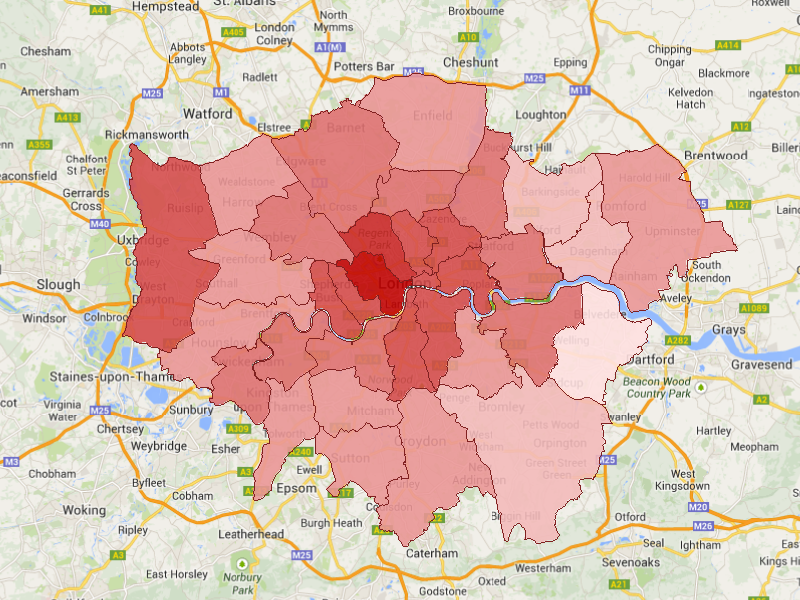}
                \small{(a) Check-ins, 26 May 2011}
        \end{minipage}%
        \hspace{0.01cm}
        \hfill
        \begin{minipage}[b]{0.32\textwidth}
                \centering
                \includegraphics[width=\textwidth]{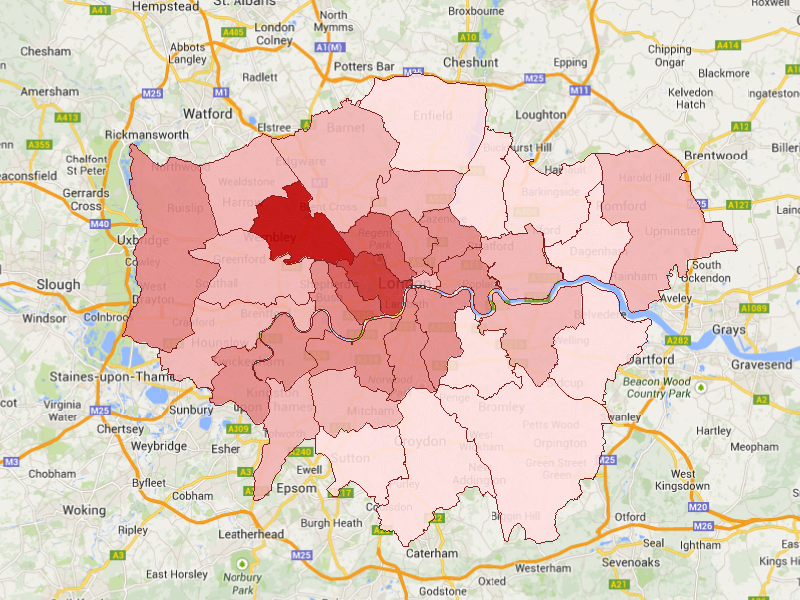}
                \small{(b) Check-ins, 28 May 2011}
                \label{fig:event}
        \end{minipage}
        \hfill
        \begin{minipage}[b]{0.32\textwidth}
              \centering
              \includegraphics[width=\textwidth]{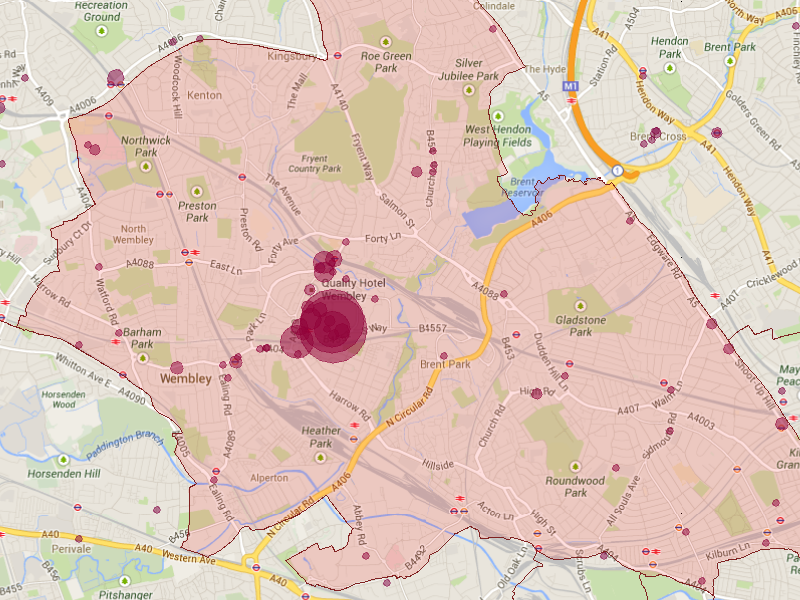}
              \small{(c) Check-ins at the Wembley area, 28 May}
              \label{fig:wembley}
        \end{minipage}%
        \caption{User check-in distribution before and during the UEFA Champions League Final event in London. Darker shaded regions denote a higher number of check-ins closer to the observed maximum among all regions during the same day. The size of the location markers in Figure (c) is proportional to the number of check-ins at the place. Notice the significantly increased activity at the Wembley area in Northwestern London on the 28th of May 2011 when the UEFA football match was held.}
        \label{fig:transition}
\end{figure*}

To reveal the underlying forces of users' attraction to certain events, we first present a methodology to mine existing events from check-in data and then test three hypotheses through which we make our major contributions:

\begin{itemize}
\item
\textbf{[H1]} \emph{Events attract users with similar mobility patterns.} To test this hypothesis we motivate the selection of core established and novel features which we subsequently evaluate in the context of the prediction task presented above. We find evidence of similarity in the past spatio-temporal activity of event participants through the hours they tend to check in at, the distance they are willing to travel, and the types of places they are inclined to visit.

\item
\textbf{[H2]} \emph{Social factors are a driving force when determining the likelihood of users to attend certain events.} Through extensive evaluation in three cities we confirm that social factors are the strongest predictors. On the one hand, event popularity, which can be related to forces of social contagion \cite{lebon2001}, dominates the results in London. On the other hand, an \emph{explicit} social filtering that checks whether friends are visiting the event tops the results in New York and Chicago, hinting at the presence of a social group identity in collective behavior \cite{aveni1977}. Third, an \emph{implicit} social signal, inspired by trust-based recommendations \cite{Jamali:2009:TRW:1557019.1557067,Andersen:2008:TRS:1367497.1367525} and based on the place type preferences of friends encoded in a socio-spatial graph, identifies relevant \emph{niche} events in London.  

\item
\textbf{[H3]} \emph{A combination of multiple factors is a more powerful signal than individual features in determining event participation preferences.} To test this hypothesis we implement a fused mobility model based on supervised learning techniques the performance of which we compare against the best single features. Overall, the prediction framework successfully identifies the exact attended event for one in three users in London and one in five users in New York and Chicago. 

\end{itemize}

Our work is one of the first to investigate event participation from the viewpoint of location-based social services. We demonstrate that such services are successful in capturing social phenomena related to crowd behavior in mass gatherings which has vital implications for personalized event recommender systems.

\section{Data Collection and Event Extraction}
\label{sec:analysis}
\begin{figure*}[!ht]
\begin{minipage}[b]{0.16\linewidth}
\centering
\includegraphics[width=2.5cm, angle=270]{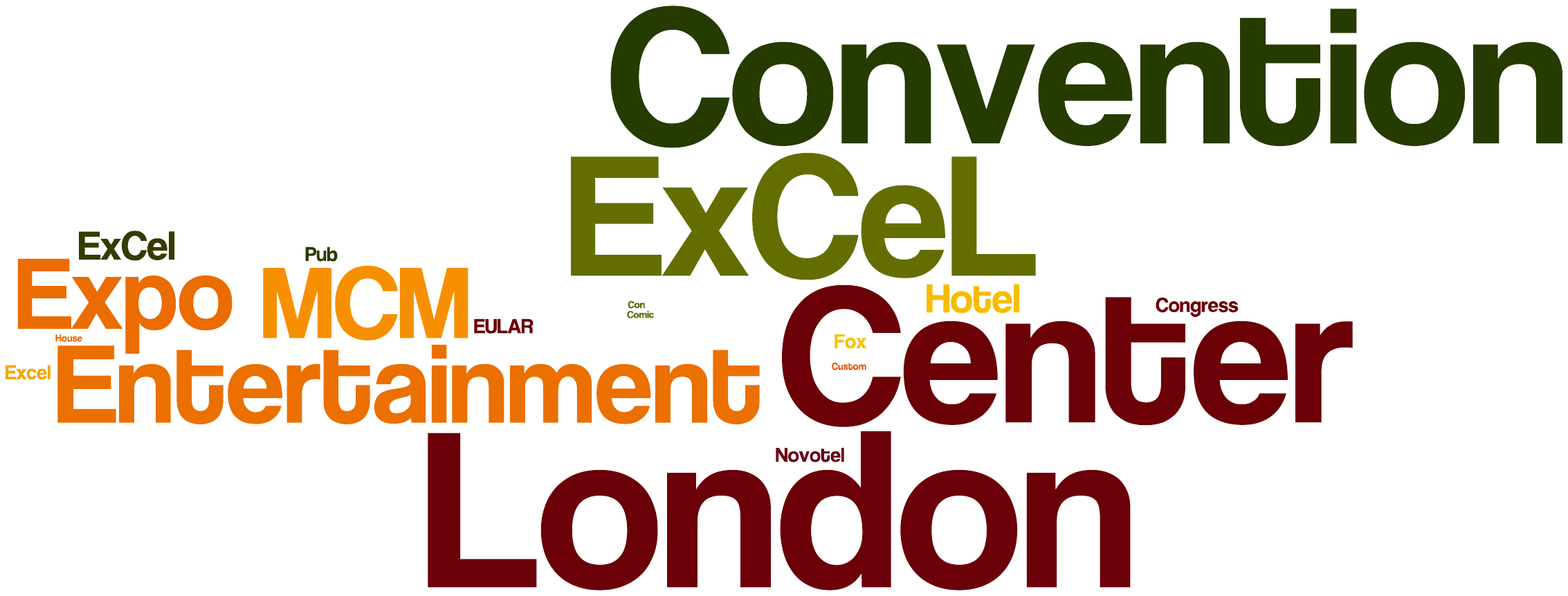}
\small{(a) MCM Expo}
\end{minipage}
\hfill
\begin{minipage}[b]{0.16\linewidth}
\centering
\includegraphics[width=2.5cm, angle=270]{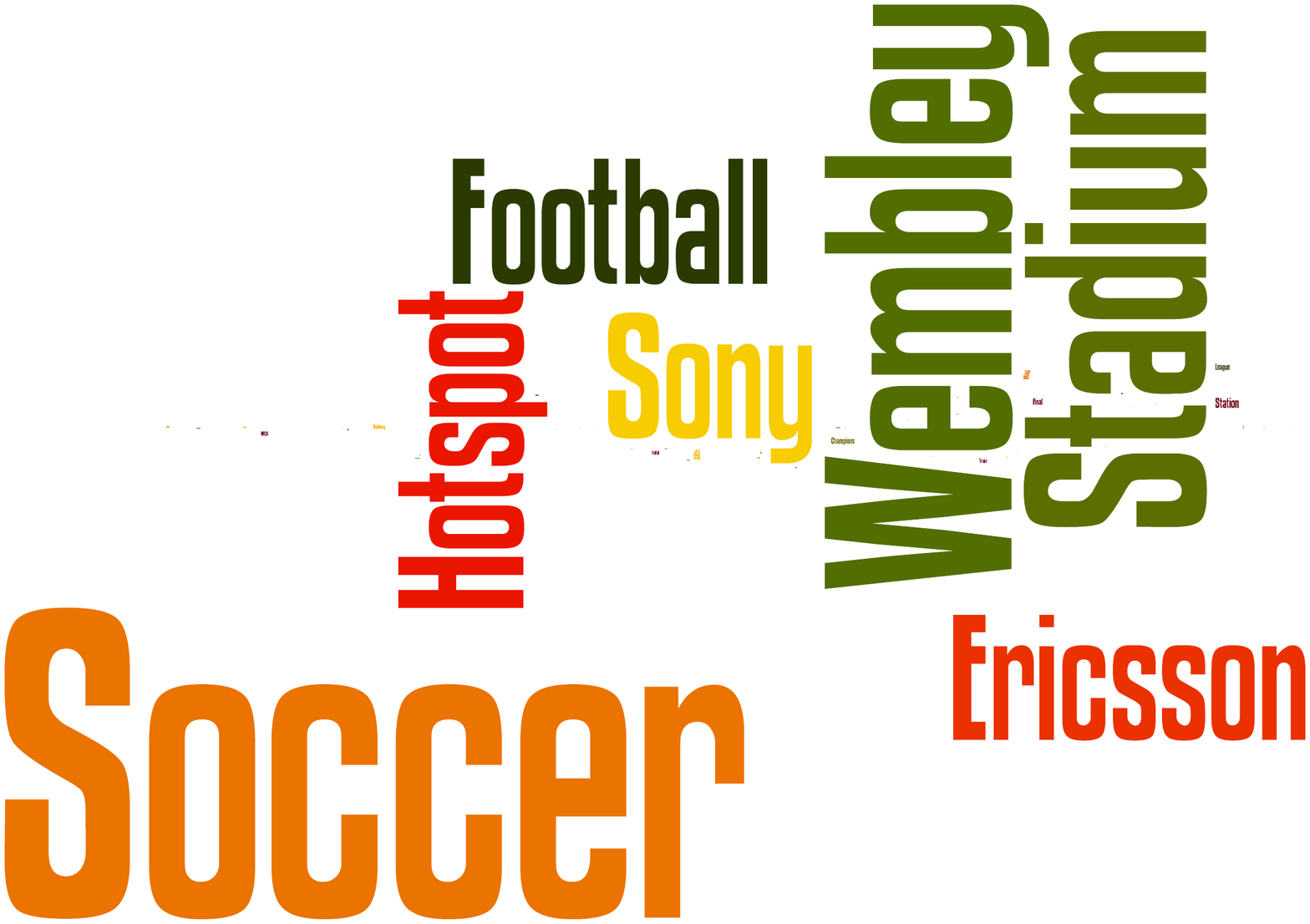}
\small{(b) UEFA Champions League Final}
\end{minipage}
\hfill
\begin{minipage}[b]{0.16\linewidth}
\centering
\includegraphics[width=2.5cm, angle=270]{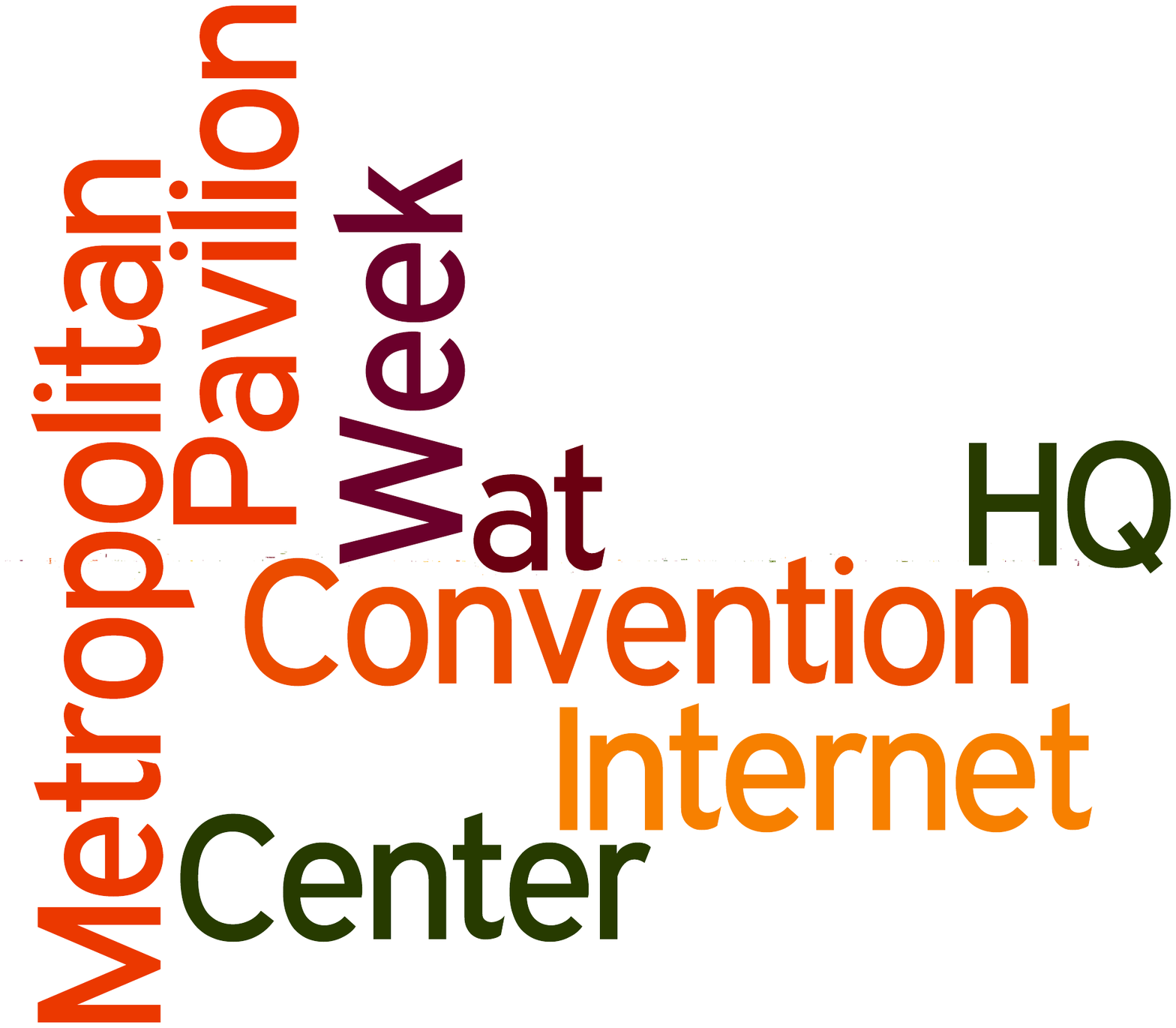}
\small{(c) Internet Week}
\end{minipage}
\hfill
\begin{minipage}[b]{0.16\linewidth}
\centering
\includegraphics[width=2.5cm, angle=270]{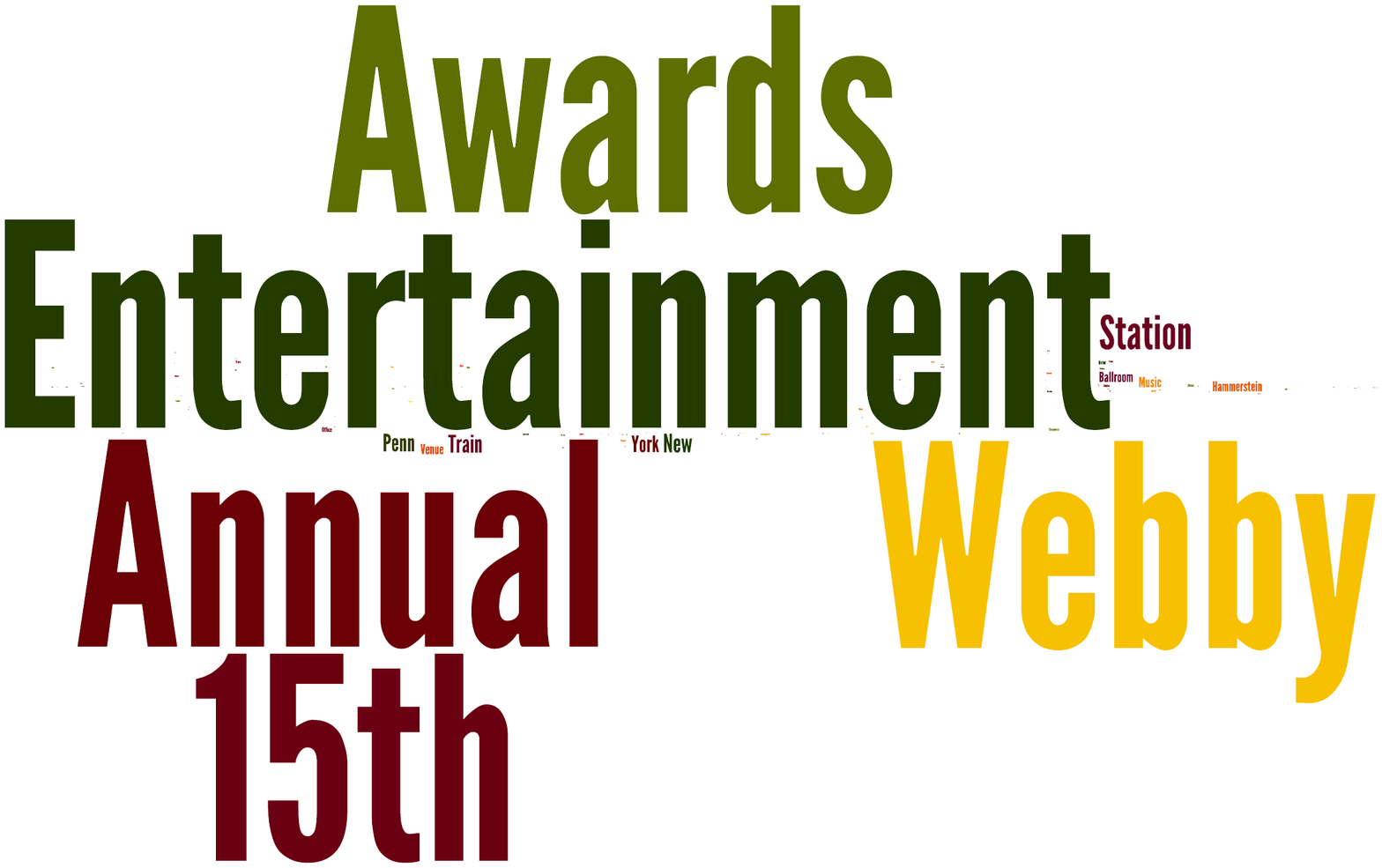}
\small{(d) Webby Awards}
\end{minipage}
\hfill
\begin{minipage}[b]{0.16\linewidth}
\centering
\includegraphics[width=2.5cm, angle=270]{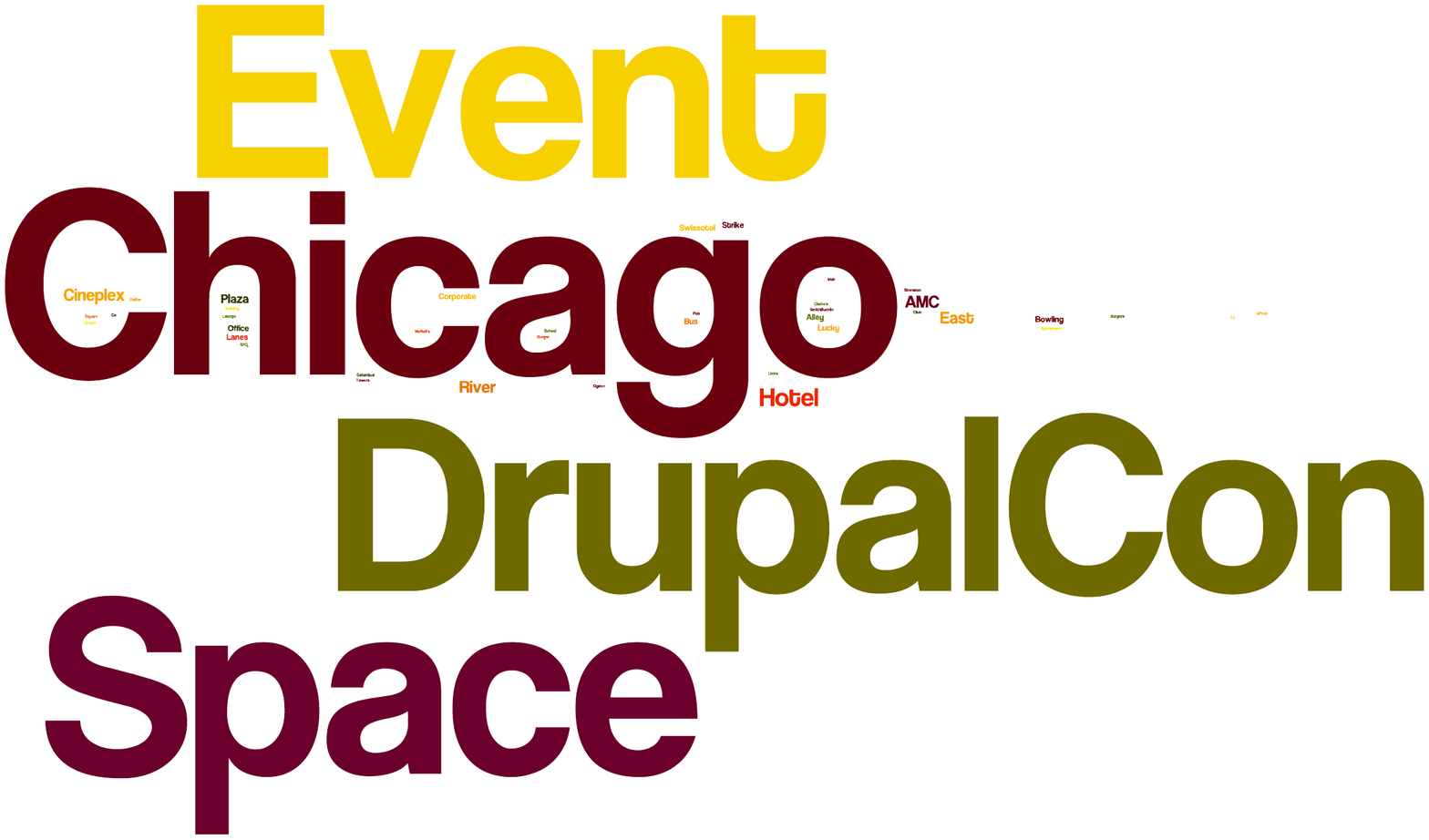}
\small{(e) DrupalCon}
\end{minipage}
\hfill
\begin{minipage}[b]{0.16\linewidth}
\centering
\includegraphics[width=2.5cm, angle=270]{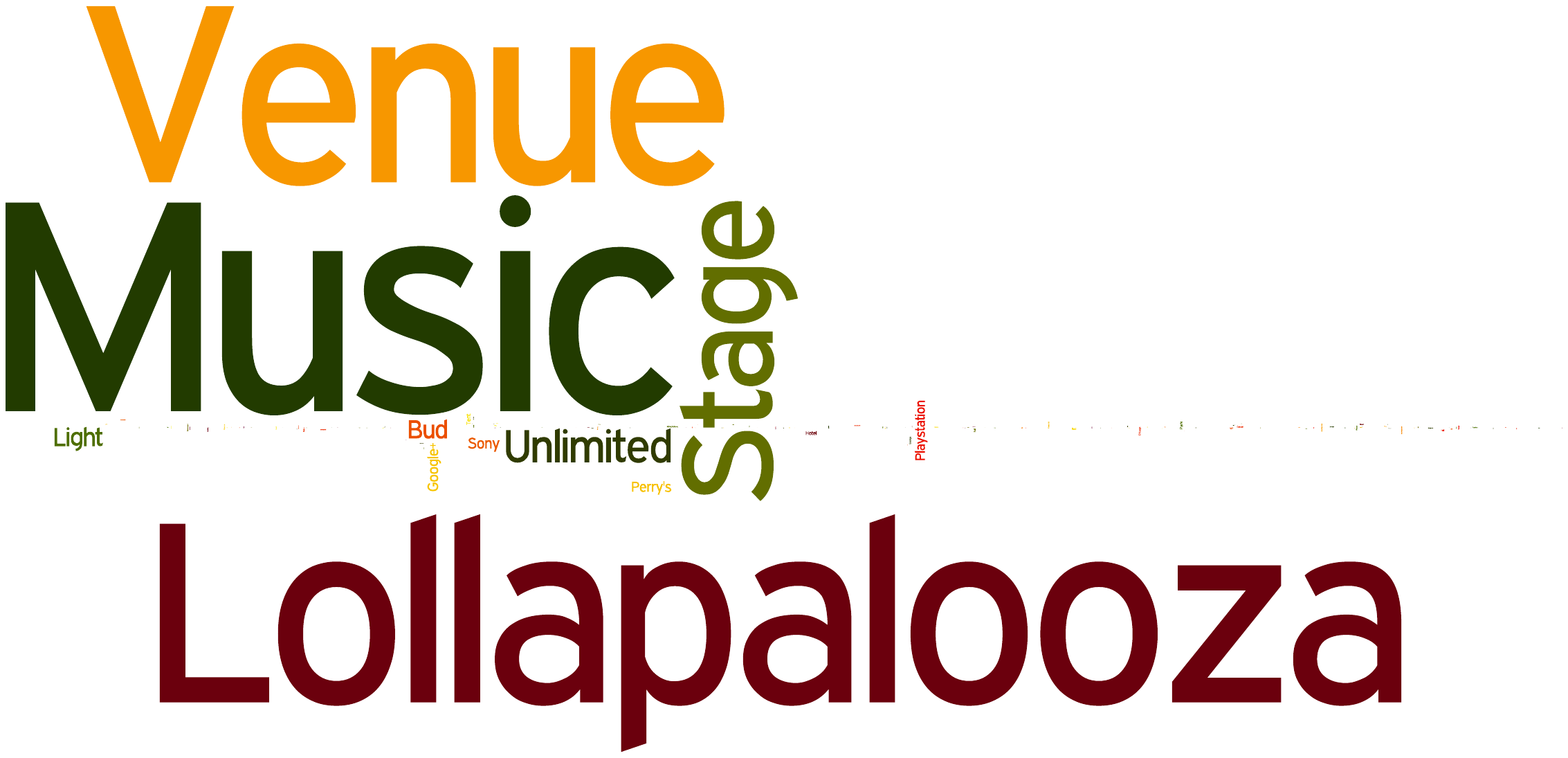}
\small{(f) Lollapalooza Festival}
\end{minipage}
\caption{Word clouds of the words used in the names of the event places and place types: (a)-(b) London, (c)-(d) New York, (e)-(f) Chicago}
\label{fig:wordles}
\end{figure*}

Foursquare, a location-based social service created in 2009, has quickly advanced as being one of the most popular location-based services with over 40 million users as of September 2013.\footnote{http://goo.gl/VNtDRP} The primary means of expressing activity through the online service is creating \emph{check-ins} which are location broadcasts tagged with tips and comments about the places visited by Foursquare users. Users can optionally share their check-ins via their Twitter accounts which enables us to crawl the check-ins via the Twitter streaming API. Over a period of 8 months, from December 2010 to September 2011, we were able to collect the 3,586,374 check-ins of 190,883 users across 184,280 venues in London, New York, and Chicago (Table \ref{table:stats}).

\begin{table}[!h]
\centering
\begin{tabular}{lrrr}
\hline
\textbf{city} & \textbf{\# users} & \textbf{\# check-ins} & \textbf{\# venues} \\ 
\hline
London & 41,397 & 533,931 & 41,701 \\ 
Chicago & 42,790 & 715,650 & 33,261 \\ 
New York & 106,696 & 2,336,793 & 109,318 \\ 
\hline
\end{tabular}
\caption{Dataset properties.}
\label{table:stats}
\end{table}

We additionally mined the city social networks from Twitter where users can subscribe to follow the public message feeds of arbitrary users. Two users are considered friends if the subscription is bidirectional, i.e. both of them follow each other's activity. Foursquare does not allow unauthorized access to a users' friend list which is the reason why the Twitter social graph is used. Although it may not be identical to the actual Foursquare graph, our evaluation results suggest it is a useful approximation sufficient for the purposes of this work. 

\subsection{Event Detection}
\label{sec:detection}
One characteristic of events is that they cause some of the places to become unusually busy on certain days. This observation has been used by Sklar et. al. in building their real-time event recommendation engine \cite{Sklar:2012:RIE:2365952.2366028} and is our guiding principle in uncovering events in the dataset. Figure \ref{fig:transition} shows the check-in attention levels of one of the most popular events in London, the UEFA Champions League Final, which was revealed by tracking changes in the popularity of the London Wembley arena during the different days. In Figure~\ref{fig:transition}b the visualized check-in levels during the event day, 28th of May 2011,  illustrate the significantly increased activity at the places in the Wembley area and hint at the simplicity of the method of tracking the place popularity for mining events.

An event is considered to be an \emph{anomalous} activity, measured in amount of check-ins, that is unusually high for a place given its check-in history. To detect events we compute the average number of check-ins per place and look for significant deviations (more than double the average) from this number during the days. The place-day pairs are then sorted in decreasing order of the absolute difference between the observed and average place popularity. For each city we pick the top 60 most popular organized events the existence of which could be verified. 

We validate the actual existence of the events at the places with increased check-in activity by performing a simple linguistic analysis on the words in the names of the event places. As shown in Figure~\ref{fig:wordles} many of the events have dedicated Foursquare places whose names exactly match the event ones. The words used to describe an event are highly informative of both its name and type (music venue, conference, football match, etc.). In the few cases when we could not obtain the exact event name from the dataset itself we resorted to manual validation via a web search engine. Although the manual labeling is a tentative task, it is a method that allows for the extraction of ground truth labels and avoids the incorporation of irrelevant items in the analysis.

\subsection{Event Scope Definition}

The dataset has a diverse set of events that may span several hours (concerts or sports games) or a whole day (festivals or conferences). To account for this diversity and not restrict the actual check-in time of users, all the check-ins at the event place that happen during the same day are considered. In addition, for some of the events we observe check-in activity at several nearby places. For instance, the UEFA finals football match has multiple check-in hotspots at the Wembley Stadium in London (Figure \ref{fig:transition}c). Once the most popular event place is identified as described in the previous section, we search for other event places with a greater than average number of check-ins in a 300-meter radius. The check-ins from these additional places are also included in the analysis. We manually verify that there are no two major unrelated events happening on the same day in the neighborhood area.

A user is assumed to have attended an event if they checked in anywhere at the event places during the day. It is possible that the true intention of users might be different from attending the event when they check in there. However, this information is not readily available and we allow for some noise in the event data.

\section{Event Participation Factors}
\label{sec:rec}
In the previous section we have extracted the check-in data for a range of events: from sports games and festivals to concerts, shows and conferences. We now pose our main hypotheses on the forces underlying users' event choices and motivate a core set of spatial, temporal and social factors.

\begin{table*}[!t]
\centering
\small
\begin{tabular}{llllllll}
\hline
\multicolumn{2}{c}{\textbf{Blogworld Expo}} & \multicolumn{2}{c}{\textbf{Orioles-Yankees Baseball}} & \multicolumn{2}{c}{\textbf{Lollapalooza}} & \multicolumn{2}{c}{\textbf{Chicago Comic Con}} \\
\hline
Place type & Score & Place type & Score & Place type & Score & Place type & Score \\
\hline
\hline
\textbf{Convention Center} & 0.0074 & \textbf{Baseball} & 0.0138 & \textbf{Music Venue} & 0.0947 & Indie Theater & 0.0106 \\
\textbf{Event Space} & 0.0033 & Bar & 0.0070 & Bar & 0.0353 & \textbf{Bookstore} & 0.0098 \\
Hotel & 0.0025 & \textbf{Sports Bar} & 0.0067 & American & 0.0195 & Convention Center & 0.0076  \\
Vegetarian / Vegan & 0.0024 & Pub & 0.0049 & Mexican & 0.0162 & \textbf{Cineplex} & 0.0072 \\
Train Station & 0.0020 & Pizza & 0.0039 & Sports Bar & 0.0162 & Other - Buildings & 0.0059 \\
American & 0.0016 & \textbf{Stadium} & 0.0038 & Pub & 0.0162 & Electronics & 0.0052\\
\textbf{Tech Startup} & 0.0015 & American & 0.0031 & \textbf{Other - Entertainment} & 0.0161 & Fast Food & 0.0047 \\
Corporate / Office & 0.0015 & Pier & 0.0030 & Corporate / Office & 0.0145 & Other - Entertainment & 0.0045 \\
Other - Entertainment & 0.0014 & Coffee Shop ~~~~~~& 0.0029 & Stadium & 0.0145 & \textbf{Movie Theater} & 0.0044 \\
Bookstore & 0.0013 & Gym & 0.0029 & Burgers & 0.0139 & Grocery Store & 0.0042 \\
\hline
\end{tabular}
\caption{Top 10 place categories observed in the check-in history of participants in four events. The first two events are held in New York, while the second two in Chicago. Place types in bold match the general theme of the event.}
\label{table:new_york_catsample}
\end{table*}

\subsection{Events and User Mobility}

The motives of visitors attending organized events can range from cultural exploration to socialization and gregariousness \cite{Crompton1997425}. Regardless of the concrete reason for participation, the event acts as a focal point for its attendees sharing a common experience. We hypothesize that \emph{some level of commonality also propagates to the mobility patterns of participants}.

\textbf{1) Attending nearby events}: Our first conjecture is that geographic distance might restrict the venue preferences of Foursquare users to nearby places and, by extension, to nearby events. Some evidence in favor of this intuition can be found in previous work suggesting that a large proportion of human movements are short-range \cite{friendshipAndMobility} and predictable \cite{BarabasiLimits}. We therefore model the role of spatial proximity by introducing the factor \textbf{\emph{Home Distance}}: a user's likelihood to attend an event is inversely proportional to the distance between their most frequently visited place, or \emph{home}, and the most popular event place.

\begin{table}[h!]
\small
\centering
\begin{tabular}{lp{7cm}}
\hline
\hline
$L$ & set of city locations \\
$C$ & set of place types \\
$U$ & set of city users \\
$E$ & set of city events \\
$L(e)$ & set of event places \\
$U(e)$ & set of event attendees \\
$G$ & city social graph \\
$G(e)$ & social network of event attendees \\
$\Gamma_{u}$ & neighborhood set of user $u \in U$ in $G$ \\
$N^c_u$ & accumulated \# check-ins for user $u$ at places of type $c$ \\
$N^h_u$ & total \# check-ins between hours $h$ and $h+1$ for user $u$ \\
\hline
\end{tabular}
\caption{Notation. In the context of a particular event and user we imply the check-ins at place category $c$ or hour $h$ up to the day before the event occurs.}
\label{table:notation}
\end{table}

\textbf{2) Place type like-mindedness}: The next dimension of event participants' potential similarity is their past activity. In Foursquare the activities, and by extension the type of past attended events, can be inferred by users' visited types of places. For instance, football matches and large concerts take place at stadiums, while festivals are typically outdoor attractions located in parks and open spaces. By looking at the types of places users tend to visit we expect to gain a broader view on the events they are interested in. 

Taking advantage of this intuition, we quantify the level of attractiveness of an event for a user by comparing the user's activity patterns to the collective activity of the event crowd. Our hypothesis is that \emph{the closer the user profile is to the collective behavior of the mass, the higher the chances are of the event attracting the user}. One way to materialize this notion through location-based data is compute the cosine similarity $cos \angle(\hat{r_u}, \hat{r_e})$ between two vectors representing the profiles of the user and the event. On the user's side, the vector $\hat{r_u}$ is built from assigning scores to the visited place types: higher values are given to categories that are popular for a particular user but at the same time are not popular among most users in general. These requirements are highly reminiscent of the Term Frequency-Inverse Document Frequency (TF-IDF) commonly used in Information Retrieval \cite{Baeza-Yates:1999:MIR:553876}. Users could be modeled as documents and the place categories as terms. The weight of a term in a document is simply the number of check-ins of a user at places of the type associated with the term. Employing the notation from Table~\ref{table:notation}, the user's score for category $i$ is defined as:
\begin{equation}
r_u^i = \frac{N^i_u}{max(\{N^j_u: j \in C\})} \times \ln{\frac{|U|}{|\{v \in U: N_v^i > 0\}|}}
\label{equation:tfidf}
\end{equation}

The aggregated event profile is similarly built from the past visited place types of its attendees where place categories are ranked differently based on their specificity for event participants. The ranking strategy should give higher ranks to place types that are common among the majority of the participants $(a)$. Higher ranks should be also given to place types whose attendance contribution from participants is relatively large compared to other place categories $(b)$. An element $r^i_e$ from the event vector $\hat{r_e}$ corresponds to places of type $i$ and is the result of the multiplication of the two factors $(a)$ and $(b)$:
\begin{equation}
r^i_e = ab =\frac{|\{u' \in U(e) : N^i_{u'} > 0\}|}{|U(e)|} \times \frac{\sum\limits_{u' \in U(e)} N^i_{u'}}{\sum\limits_{u' \in U} N^i_{u'}}
\label{equation:catscore}
\end{equation}

We call this metric that captures the "herding" behavior of participants the \textbf{\emph{Place Category Score}}. When building the event profiles and looking at the related place types, we see that the adopted metric is highly effective in uncovering an important aspect of event attendance preference. As demonstrated in Table~\ref{table:new_york_catsample} where the top 10 most highly ranked place categories for events are listed, participants in an event appear to have a preference to visit places of a similar type as the one of the most popular/central event place. In our dataset, baseball and football matches, for instance, attract fans that previously visited Stadiums, conferences appear to attract people visiting Convention Centers, and concerts attract users visiting Music Venues.

\begin{figure}[htp]
		\centering
		\includegraphics[width=0.46\textwidth]{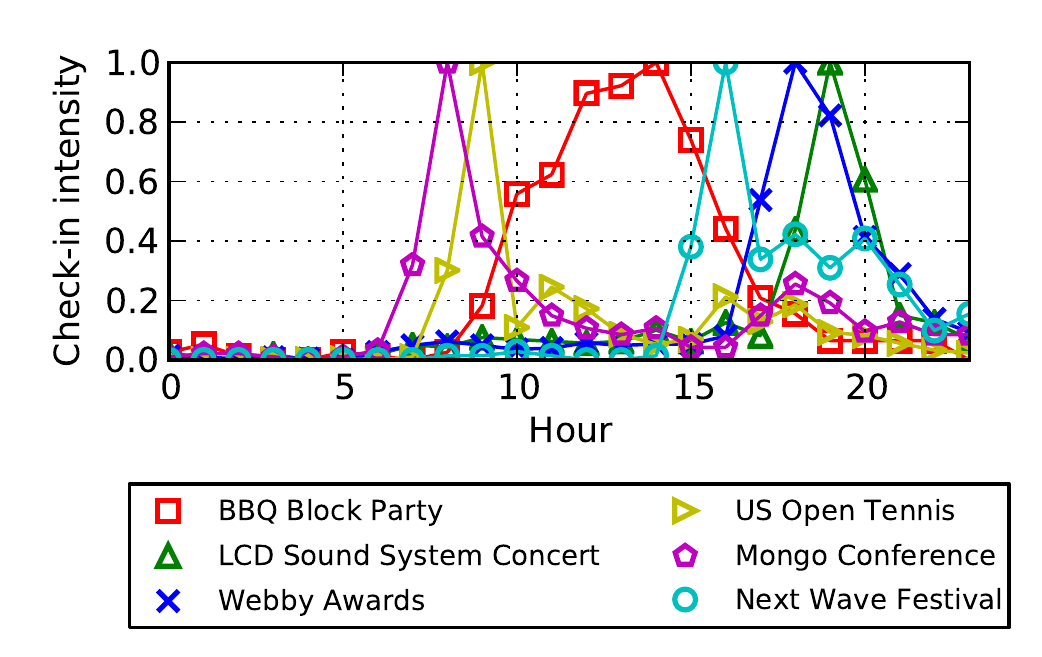}
        \caption{Temporal shapes of six events in New York city. The number of check-ins
        at each hour is normalized by the maximum value reached on the day of the event. A global maximum, or a peak, is often observed.}
        \label{fig:shapes}
\end{figure}

\textbf{3) Hourly patterns:} A third dimension of the factors driving users' decision to visit events is the temporal preferences of users to get involved in activities. Our assumption is that \emph{if a user is mostly active during a particular time of the day such as the evening, they would rather attend an event aligned with these temporal preferences}. As an approximation for the event time we could adopt its peak hour. As portrayed in Figure~\ref{fig:shapes}, the temporal distribution of check-ins at events of different types usually has a well-defined shape that reflects how users arrive at the event venues before and during the event. The peak is an often observed phenomenon that marks the onset of an expected activity such as the beginning of a concert or a sports game. In fact, more than 40\% of the events in all cities have at least half of their check-ins created at the peak and the hours immediately before and after it. The alignment between the event and the user's past temporal activity can then be captured by measuring the extent to which users tend to check in at the hours around the event peak $\hat{p}_e$: 
\begin{equation}
\hat{d} = \sum_{h \in [0, 24)}{\frac{N^h_u}{\max{N^h_u}} \times min(|h - \hat{p}_e|, 24 - |h - \hat{p}_e|)}
\end{equation} 
A small \textbf{\emph{Temporal Distance}} $\hat{d}$ implies that the user prefers to check in predominantly at hours coinciding with the event ones.

\subsection{Events and Social Forces}

Our next main hypothesis is \emph{that social forces are the primary means of luring users to attend events}. These forces can take various forms and in this section we give three prominent examples.

\textbf{4) Following the crowd}: A strong motivation for users to participate in an event might be its \textbf{\emph{Popularity}}, which can be measured in amount of check-ins. Attending events because of their popularity can be considered as a form of \emph{crowd behavior} where individuals follow trends through social \emph{contagion} or \emph{imitation} \cite{lebon2001}. This claim is partly supported by our findings that the events in the dataset feature a long-tail popularity distribution. A few of them attract large masses of users (such as the Royal Wedding in London or Lollapalooza Festival in Chicago), while the rest have a markedly lower number of users checking in at related venues.

\textbf{5) Social group identity:} Many events such as festivals and concerts are social activities by nature which is why we expect the purely social motivation for users to attend certain events to be particularly strong. This intuition is confirmed in social systems with respect to crowd behavior \cite{aveni1977} where participants in mass gatherings are more likely to be found among a group of friends. Drury and Reicher \cite{reicher1999} further develop a social identity model by proposing that crowd behavior is driven by intergroup dynamics where individuals adopt the collective identity of their social group to interact with others. Falling back on these studies, we put forward a \textbf{\emph{Social Influence}} factor that assumes that \emph{a user would prefer to join events for which the number of visiting friends is larger}. 

It is possible that some events are ranked equally high for a user because the number of friends attending is the same. We argue that in this case the probability of joining the events may not be the same and it often depends on the social importance of the event for the user's friends. In such situations we break ties by considering the maximum degree centrality ($|\Gamma_{u} \cap U(e)|$) of a friend in the social network of event attendees. Our reasoning is that if an event is of a particular interest to a friend, they would most probably play a central role in the social network of attendees and would attract more of their friends in turn to participate.

\label{sec:rw}
\textbf{6) Place-focused social interactions:} Our next hypothesis is that \emph{the friends' visited place categories and the associated activities with them can be indicative of the users' event preferences}. The types of places visited by friends may act as gravitational forces for social interactions where friendship is fostered and ultimately manifested in collective participation in events. This could be considered as a type of homophily in social systems \cite{mcpherson2001birds} where networks are homogeneous with respect to behavioral characteristics. In our case the homogeneity is captured though the common place types such as Bars and Theaters where friends meet.

\begin{figure}[htp]
\centering
\includegraphics[width=0.30\textwidth]{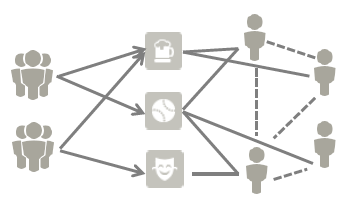}
\caption{An example socio-spatial graph. Nodes represent events (left), users (right) and place types (middle). Dashed links denote social network relations. User-user and user-category links are bidirectional but have different weights depending on the direction. A random walk starts from an event node and reaches out to users via place types.}
\label{fig:rw_example}
\end{figure}

To model the above mentioned assumptions we design a graph that seamlessly combines social and spatial signals and that connects users, place types and events as shown in Figure~\ref{fig:rw_example}. Personalized random walks with restart \cite{fastRWR} are performed on the graph to compute user attraction scores towards events. 

\textbf{Graph Definition}.
The graph is a directed one with three types of nodes: users $U$, events $E$ and place categories $C$. There are three types of links which we weigh differently to encode domain-specific transition probabilities. \emph{User-user} links connect two users $i$ and $j$ if they are friends. The weight $w_{ij} = \frac{1}{|\Gamma_{i}|}$ of the connecting arc is inversely proportional to the number of friends user $i$ has. \emph{User-category} links connect users to their visited place categories. The arc between user $u$ and category $c$ is weighed according to the TF-IDF score as defined in Equation \ref{equation:tfidf}. The reverse link is weighed similarly with the difference being that now a place category is represented as a document and the users checking in there as terms: 
\begin{equation}
w_{cu} = \frac{N^c_u}{max(\{N^c_v: v \in U\})} \times \ln{\frac{|C|}{|\{j \in C: N_u^j > 0\}|}}
\end{equation}
Last, an \emph{event-category} link connects an event to a place type if it is among the top $K$ categories visited by users participating in the event. The place types are sorted in descending order of the place category score (Equation (\ref{equation:catscore})) which is used as the weight on the links. A too low value of $K$ might overlook an important place preference signal, while a too high value might introduce unwanted noise. We find that $K = 10$ offers a good balance between sufficient detail and tolerable noise. Finally, to correctly set transition probabilities in the resulting graph, we normalize the weight on each link by dividing its value by the sum of the weights on the out links of the source node.

\textbf{Random Walks with Restart}.
Random walks on graphs have been used to rank nodes in a way that encodes the probability of reaching a target node from a source. The ranking information has been successfully employed by variations on the PageRank algorithm \cite{pagerank} to compute importance scores of web pages in the web page citation graph.
Random walks with restart \cite{fastRWR} are personalized versions of the model that additionally incorporate a constant probability of jumping back to a specific graph node in order to bias the walks nearer the node's neighborhood. The restart step is essential for acquiring a personalized view of the graph with respect to a specific node. In our case, this allows us to measure the extent to which a user is related to a concrete event.

A random walker starts from an event node, keeps traversing adjacent links and with constant probability $(1-\alpha)$ jumps back to the event node which guarantees the personalized view of the graph. The parameter $\alpha$ is a scaling factor that is usually set to 0.85 \cite{pagerank}. By setting the restart probabilities at other nodes to zero we ensure that the random walker explores nodes close to the event neighborhood more often. We are then interested in the steady-state probability that we reach the user nodes. If a user is easily reachable from an event via place types, friends or any combination of factors, the random walk score of the user node in the graph will be higher. The preference towards events is considered stronger when the computed user random walk scores are higher.

\section{Experimental Evaluation}
\label{sec:eval}
In this section we formulate an event prediction task in the context of which we evaluate the strength of the described factors. In doing so we confirm our hypotheses: [H1] there is similarity in the event participants' mobility patterns; [H2] social signals and popularity play a leading role in the prediction task; [H3] a combination of factors is more informative of users' event preferences compared to individual features.

\subsection{Evaluation Methodology}

We define an event participation prediction problem as follows: given a set of events and a set of users, find a way to rank events so that those at the forefront of the prediction list are the ones that the user actually attends. Events are ranked according to the preference scores produced by the participation factors as described in the previous section. In this context the factors behave as prediction features. A feature is considered more successful in explaining the motivation behind a user's participation if it gives higher ranks for events that are truly attended by the user.

For the evaluation, we use a stratified $10$-fold cross validation with respect to users. From each event $10$\% of the participants are repeatedly held out as test users. The rest of the users in the training set are assumed to be the ones who have signed up for the event and they are the ones from whom the event profiles are built. 
When building the user and event profiles, only the check-in activity prior to the day of the event is considered without including the check-in data from the event itself. For each test user all items are ranked and a single preference list of events is produced many of which happen in different days.

\subsection{Metrics}

The performance of the event ranking features is evaluated with respect to two metrics: normalized discounted cumulative gain (NDCG) and accuracy. The $NDCG@N$ metric is commonly used in information retrieval \cite{ndcg} to measure the effectiveness in the ranking of relevant items in a list of recommendations:
\begin{equation}
NDCG@N = \frac{1}{Z_N} \sum_{i=1}^{N}{\frac{2^{rel(l_i)} - 1}{log_2(1 + i)}}
\end{equation}
The relevance $rel(l_i)$ of an item (event) $l_i$ at position $i$ in our case is equal to 1 when the user attended the event and 0 otherwise. The idealized cumulative gain $Z_N$ is a normalizing constant such that a perfect ranking with all relevant items ordered first would result in an NDCG value of 1. 
We also use the Accuracy@N metric which for a user is defined as 1 if and only if an event that the user attended is ranked within the top $N$ items in the prediction list. The accuracy results are averaged across users. This metric is complementary to the NDCG one and shows for what proportion of the users a feature brings relevant events to the front of the prediction list. The Accuracy@X\% is similar and represents the cut-off threshold equal to X\% of the total number of events eligible for prediction.

\begin{figure*}[!ht]
\centering
\includegraphics[width=\textwidth]{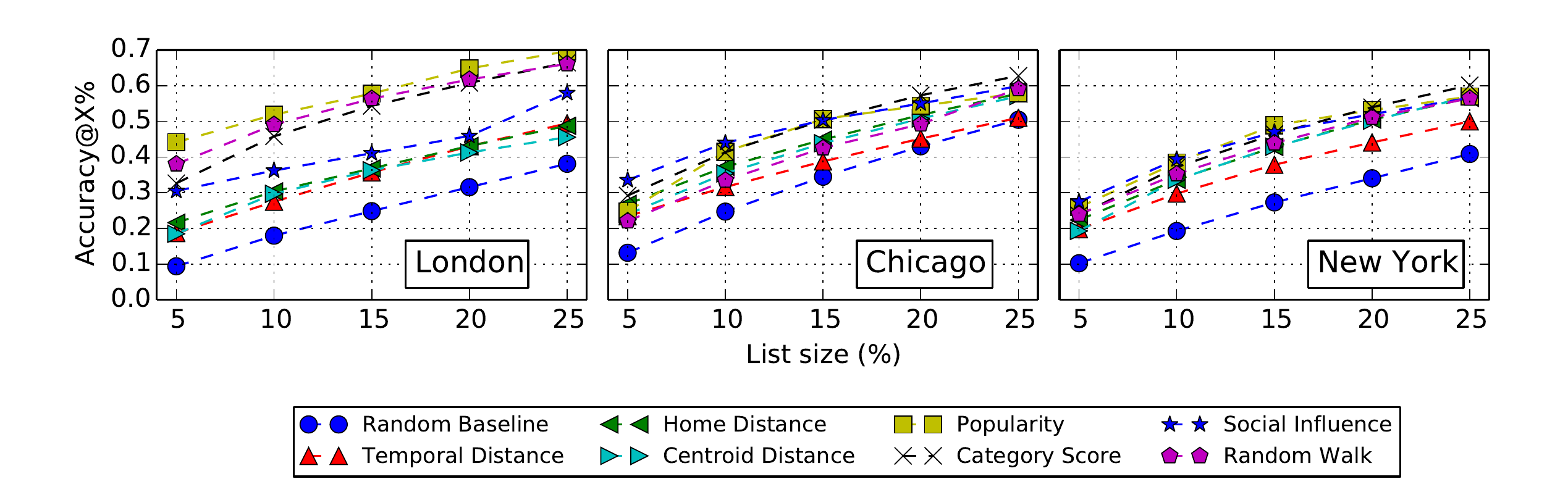}
\caption{Averaged user accuracy of the mobility features as a function of the prediction list size.}
\label{fig:accuracy}
\end{figure*}

\begin{table}[h!]
\small
\centering
\begin{tabular}{cccc}
\hline
\textbf{Model} & \textbf{London} & \textbf{Chicago} & \textbf{New York} \\
\hline
\hline
\textit{Random} & \textit{0.118} & \textit{0.142} & \textit{0.115} \\
Temporal Dist. & 0.203 & 0.221 & 0.194 \\
Home Dist. & 0.219 & 0.245 & 0.223 \\
Category Score & 0.315 & 0.267 & 0.235 \\
Popularity & \textbf{0.411} & \textbf{0.275} & \textbf{0.262} \\
Social Influence & 0.290 & \textbf{0.306} & \textbf{0.268} \\
Random Walk & \textbf{0.347} & 0.221 & 0.244 \\
\hline
\end{tabular}
\caption{Averaged \emph{NDCG@10} for the different ranking strategies. Top 2 features for each city are in bold.}
\label{table:ndcg}
\end{table}

\subsection{[H1]~Events and User Mobility}
Here we test our hypothesis that event interests imply some similarity in the mobility patterns of participants. We confirm this by comparing the spatio-temporal features' performance against a random baseline (Table \ref{table:ndcg}). The \emph{Temporal Distance} and \emph{Home Distance}, albeit being weaker than the other signals, still perform significantly better than random, which implies similarity in the temporal and spatial dimensions: 1) events are likely to attract users that historically check in more at hours around the event peak, and 2) events appeal more to users that are geographically close to the activity hot spot. Among the spatial-only factors, the one that encodes the \emph{Place Category} preferences of users performs better than the simple distance-based metric in all cases. This confirms that the semantics (types) of places are more informative than pure distance when it comes to event preferences, which is unlike standard place mobility models where distance is a dominant factor \cite{scellato2011socio}. These observations suggest that place types alone, as already hinted by the \textit{Random Walk} model that incorporates them too, can be an important source of information for inferring event interests since certain events appear to attract users with common activity patterns captured in check-ins at particular types of places.

\subsection{[H2]~Events and Social Forces}
\textbf{Friends in the Crowd:} In this part of the analysis we test whether social factors in their various forms are driving forces for event participation. A first discovery in testing this claim is that event popularity, as captured by the number of attendees, is truly among the best predictors across cities. In London the feature achieves the highest NDCG score, $0.411$, observed for a factor. The reason for this is that there are massively popular events in cities, such as the Royal Wedding and the UEFA Champions League Final in London, that attract a large number of people. This phenomenon is highly reminiscent of preferential attachment models \cite{Albert2002} where popular entities (events) lure even more followers governed by forces such as gregariousness and social contagion. 

On the other hand, we confirm that the \emph{Social Influence} feature is extremely strong in the event domain, scoring as high as $0.306$ in Chicago and $0.268$ in New York and outperforming even \emph{Popularity} which reaches $0.275$ and $0.262$ in the two cities respectively. These figures suggest that events foster social participation which is in line with Aveni's findings on the role of social groups in collective behavior \cite{aveni1977}. We recall that when designing the \emph{Social Influence} feature we additionally incorporated the degree centrality of the user's most socially involved friend as a way to measure the social importance of an event for a user. To understand whether this additional complexity is worthwhile, in Table ~\ref{table:social_filt} we compare the performance of the enhanced signal to the no-centrality baseline for users that have events with an equal number of participating friends. The significant, more than $12$\% improvement in \emph{Accuracy@1} implies that the centrality technique is successful in breaking ties among already highly ranked events. This suggests that the preference towards an event for a user can be successfully inferred based on the social engagement of their friends. 

\begin{table}[h!]
\small
\centering
\begin{tabular}{lrrrrr}
\hline
\multicolumn{2}{c}{} & \multicolumn{2}{c}{\textbf{NDCG@10}} & \multicolumn{2}{c}{\textbf{ACC@1}} \\
\hline
City & \# Users & Base & Centr. & Base & Centr. \\
\hline
\hline
London & 843 & 0.43 & 0.46 & 0.32 & 0.36  \\
Chicago & 2323 & 0.31 & 0.33 & 0.21 & 0.26 \\
New York & 3972 & 0.38 & 0.40 & 0.24 & 0.28  \\
\hline
\end{tabular}
\caption{Comparison of the \emph{Social Influence} feature performance between its two variants: with and without centrality. These results are obtained through leave-one-out cross validation and averaged across users for whom there are at least 2 events with an equal number of friends.}
\label{table:social_filt}
\end{table}

In terms of Accuracy@X, the results shown in Figure~\ref{fig:accuracy} are consistent with the \emph{NDCG} ones: the best performing features are the socially influenced models and in the case of London, \emph{Popularity}. For all cities the Accuracy@5\% for the \emph{Social Influence} is around 30\% which means that roughly for 1 in 3 users on average the metric correctly identifies a relevant event within the top 5\% of the items in the prediction list. This hints that social factors are better at predicting the exact event a user would attend as discussed above.

\label{sec:niche}
\textbf{Where Friends Meet:} 
An intriguing outcome is that the implicit social signal hidden in the place type preferences of friends and captured by the \emph{Random Walk} model exhibits diversity in its performance (Table~\ref{table:ndcg}). In London it achieves a high score of $0.347$ ranking second best overall, whereas in Chicago and New York the results of $0.221$ and $0.244$ respectively are clearly lower than the ones of \textit{Social Influence} and \textit{Popularity}. We demonstrate that this heterogeneity is related to the presence of \emph{niche} events that engage users who prefer to check-in at place types that are not generally popular.

A niche event such as a football game can be characterized by the highly targeted interests of its fans. This can be reflected in the participants' place type preferences where visiting certain place categories such as football stadiums may be common among the attendees, or among certain friend circles, but not popular in general. To formalize the notion we look at the Kendall's $\tau$ correlation coefficient \cite{kendall1938measure} between the ranking of place categories for an event profile, as shown in Table~\ref{table:new_york_catsample}, and the overall ranking of the place type popularity as reflected in the Foursquare data. An event is considered more \emph{niche} if its Kendall's $\tau$ correlation coefficient is lower or negative. In such cases the discrepancy in the two rankings dominates which implies that there are place types less popular among the common user but high on the list among event attendees.

\begin{table}[h!]
\small
\centering
\begin{tabular}{ccc}
\hline
\textbf{London} & \textbf{Chicago} & \textbf{New York} \\
\hline
$-0.50^*$ & $-0.38^*$ & $-0.42^*$ \\
\hline
\end{tabular}
\caption{Spearman correlation coefficients between the Kendall's $\tau$ score and the Accuracy@5\% of the random walk model, p-value  $<0.01^*$.}
\label{table:correlations}
\end{table}

The key observation illustrated in Table~\ref{table:correlations} is that there is a statistically significant negative correlation between the Kendall's $\tau$ score and the Accuracy@5\% for the random walk model on the socio-spatial graph. This means that the more niche an event is, the better the performance of the random walk model becomes. In London 62\% of the events have a negative Kendall's $\tau$ score implying a highly niche content for their participants. In contrast, there are only three such events in Chicago and zero in New York. Our findings recognize \emph{the influence of friends and common interests on the motivation to visit niche or special events of value which is higher to the social group than to the general community}.

\label{sec:supervised}
\subsection{[H3]~Inter-signal Interactions}

We have observed that while certain features such as \emph{Social Influence} and \emph{Popularity} dominate in most cases, there is some heterogeneity in their relative performance across cities. We have also seen that the random walk model performs well for niche events which can vary in number from one city to another. The question we address here 
is whether we could adopt a supervised learning procedure for combining participation features into a fused prediction system that automatically dissects the heterogeneities and outperforms the individual participation factors.
By building this framework we hypothesize that a combination of factors better reflects users' decision to participate in an event.

\textbf{Training Strategy.}
The features we have examined produce a score for a user-event pair which indicates the likelihood that a user attends an event. For each user-event example we build an instance by assembling the scores of the individual predictors into a feature list and appending a positive (+1) or a negative (-1) label depending on whether the user truly attended the event. A training set is built from a subset of the users. For each user we include the positive examples as well as 15 randomly chosen instances corresponding to events the user has not attended. Regression models are trained that produce a real-valued output for user-event pairs which allows us to rank events according to the predicted preference scores.

\textbf{Evaluation Strategy.}
We adopt the same $10$-fold cross validation procedure as presented in the "Evaluation Methodology" Section. The difference is that now for each of our training users we have a set of positive and negative examples which constitute the training set. Note that although the training phase includes a reduced set of user-event samples, in the testing phase we evaluate against all possible combinations of test users and events. The supervised learning algorithms we have experimented with are linear ridge regression (with the regularization parameter being set to $\lambda$=$10^{-8}$)~\cite{Hoerl1} and M5 model trees \cite{Quinlan92learningwith}. We have used the publicly available implementations in the WEKA framework \cite{Weka}. Two versions of the algorithms are considered: one that combines all features and one that excludes the random walk probability scores from the socio-spatial graph. This separation allows us to evaluate an additional hypothesis: \emph{the place type preferences of friends implicitly expressed with the random walk scores are a fundamentally different signal not captured in a combination of other features}.

\textbf{Results.}
By comparing the supervised models against the single predictors in Table~\ref{table:supervised}, we find that the M5+RWR trees attain the best performance. They outperform the incorporated best single features by a clear margin ($0.117$ for London, $0.057$ for Chicago, and $0.099$ for New York) and better the results of the linear regression models. This suggests that a combination of temporal, spatial and social signals integrated into a supervised learning framework can prove highly effective in predicting the participation of users to events in location-based services. 
Furthermore, the regularized linear regression model does not provide consistently good results, even when it is compared with the single features. As in the case of Chicago, the linear regression classifier LR achieves a score of $0.311$ which is only slightly above the $0.306$ value of the \emph{Social Influence} feature. Thus, a non-linear combination of features may provide a more effective modeling recipe in inferring the event interests of Foursquare users. A similar finding with respect to non-linearly mixing spatio-temporal signals for personalized venue search in Foursquare has been highlighted by Shaw et al. \cite{Shaw:2013:LRS:2433396.2433485}.

\begin{table}[htp]
\small
\centering
\begin{tabular}{cccc}
\hline
\textbf{Model} & \textbf{London} & \textbf{Chicago} & \textbf{New York} \\
\hline
\hline
\textit{Random} & \textit{0.118} & \textit{0.142} & \textit{0.115} \\
Popularity & 0.411 & 0.275 & 0.262 \\
Social Influence & 0.290 & 0.306 & 0.268 \\
\hline
LR & 0.481 & 0.311 & 0.336 \\
M5 & 0.494 & 0.346 &  0.344 \\
LR + RWR & 0.505 & 0.324 & 0.343 \\
M5 + RWR & \textbf{0.528} & \textbf{0.363} &  \textbf{0.367} \\\hline
\end{tabular}
\caption{Averaged \emph{NDCG@10} for the different supervised learning algorithms.}
\label{table:supervised}
\end{table}

\begin{table}[h!]
\small
\centering
\begin{tabular}{cccc}
\hline
\textbf{Model} & \textbf{London} & \textbf{Chicago} & \textbf{New York} \\
\hline
\hline
\textit{Random} & \textit{0.037} & \textit{0.051} & \textit{0.036} \\
Popularity & 0.267 & 0.168 & 0.151 \\
Social Influence & 0.220 & 0.198 & 0.160 \\
\hline
LR & 0.293 & 0.152 & 0.179 \\
M5 & 0.344 & 0.205 &  0.185 \\
LR + RWR & 0.307 & 0.165 & 0.182 \\
M5 + RWR & \textbf{0.372} & \textbf{0.229} &  \textbf{0.212} \\\hline
\end{tabular}
\caption{Averaged user \emph{Accuracy@1} for the different supervised learning algorithms compared against \emph{Popularity} and \emph{Social Influence}.}
\label{table:supervised_accuracy}
\end{table}

Our further hypothesis is that the place type preferences of friends are a fundamentally different signal not captured as a combination of other features. We confirm this by the important observation that using the random walk scores as a feature in the supervised learning framework improves the results for both the linear regression and the M5 model trees algorithms. In Chicago, for instance, the averaged $NDCG@10$ for the random walk on the socio-spatial graph achieves a score of 0.221 which is lower than the home distance. When this random walk signal is fused into the M5 tree, the results soar to 0.363 which is much higher than the 0.306 value of the best performing feature. Similar outstanding results are valid for London and New York as well.

In terms of user accuracy it is also notable that the only model that is able to substantially outperform the \emph{Accuracy@1} of the best single feature across all cities is the M5 Tree + RWR (Table ~\ref{table:supervised_accuracy}). The accuracy for London goes as high as 37\% which means that roughly for 1 in every 3 users the model correctly identifies the \emph{exact} event the user will attend. 
Given the results, the supervised framework accurately identifies the preferred event for one in three users in London and one in five users for New York and Chicago.

\section{Discussion and Implications}
\label{sec:implications}
The analysis and subsequent evaluation of the event participation prediction problem in Foursquare has revealed interesting insights both on the \textit{nature} of social events, as seen through the lens of location-based services, and the algorithmic strategies one may employ to recommend events.

The superiority in the performance of social signals can be eventually identified on three fronts. First, event popularity, which can be related to the strong social urge to follow trending behavior, is topping the results in London. Second, the \textit{explicit} social filtering which accounts whether friends are attending an event has performed very well in all cities. To some extent, this behavior could be attributed to the presence of a \emph{social identity} where individuals participate in the event to share collective experiences with friends. Third, the mechanics and performance of the random walk strategy have uncovered the presence of an \textit{implicit} social signal hidden in the user preferences (interests) for particular place types and by extension to specific event types. This could be viewed as a form of \textit{homophily} that brings together like minded users to social events. We have shown that in the cases of \emph{niche} events this signal yields excellent performance.

Although we have observed some diversity in the performance of the various participation features both across event types and cities, we have offered a recipe that copes with these issues. A supervised learning approach has proven effective in combining the different information signals into a unified framework so as to provide top performance in all contexts. In the event recommendation task, our findings suggest not only that combining multiple factors is highly desirable, but also that extracting social signals is of utmost importance for achieving high accuracy. 

These results should be interpreted in the context of potential biases originating from the data collection and the check-in process in Foursquare. On the one hand, our dataset relies on users who have explicitly shared their whereabouts via Twitter. According to Scellato et al. \cite{scellato2011socio} such users constitute between $20\%$ and $25\%$ of the total Foursquare population in 2010. On the other hand, it is hard to validate the true intention of the users when they check in at particular venues. As we primarily focus on studying aggregated behavior from a large user base, however, our approach is able to tolerate a certain amount of noise.

\section{Related Work}
\label{sec:related}
\textbf{Event Mobility Analysis and Detection}. 
To our knowledge, event analysis so far has been limited to isolated cases and specific types of events. Xavier et al. \cite{largeScaleEvents}, for instance, focus on mobility aspects of users during large-scale events but fail to provide any insights as to why users attend the event. Calabrese et. al. \cite{Calabrese:2010:GTA:2166616.2166619} have studied crowd mobility during special events but they have solely concentrated on correlating the type of the event with the origin of people attending it.
Only recently have online social networks entered the event detection arena \cite{sakaki10} due to the massive amounts of timely user-generated content in response to external anomalous events. Sklar et al. \cite{Sklar:2012:RIE:2365952.2366028} have built a real-time event detection engine in Foursquare that is based on a probabilistic model for measuring how unusually busy a place becomes. Although they recommend the detected nearby events to users, they do not focus on understanding the relationship between the user past check-in patterns and the likelihood of attending certain events.

\textbf{Event Prediction}. 
The event prediction problem has been studied by Quercia et al.~\cite{Quercia:2010:RSE:1933307.1934616} when providing cold-start event recommendations for users whose home location is known. However, the authors have not focused on personalization. Three other prominent examples of event recommender systems have been built in the domains of on-going cultural events, scientific and conference talks. Lee~\cite{Lee:2008:PTC:1454008.1454060} exploits trust relations together with explicit user feedback to recommend cultural events, while Minkov et al. \cite{Minkov:2010:CFE:1871437.1871542} combine content-based with collaborative filtering approaches to capture user preferences towards latent topics hidden in scientific talk announcements. Liao et al.~\cite{Latent} further develop latent models based on offline spontaneous interactions and co-attendance information to recommend related events in offline ephemeral social networks formed around conference talks. In comparison to these works, the events that we study in location-based social services currently lack many of the contextual advantages that the above mentioned systems take for granted: explicit event preference information, on-going nature of specific events, detailed topic descriptions and offline interaction data.

\section{Conclusions}
\label{sec:conclusion}
In this work we have studied the spatio-temporal and social forces behind users' decisions to attend certain events as seen through location-based social networks. We have defined a prediction framework that at the expense of some potential attendance knowledge assesses different dimensions of homophily effects observed through collective participation in events. While social forces tend to dominate over the others, confirming theories on crowd behavior, we uncover some heterogeneities in the performance of the prediction features across cities and event types. This proves that combining the disparate signals into a supervised learning framework for event participation prediction is necessary for obtaining top performance in all cases. The insights drawn and the framework developed in this work could help towards designing better personalized event recommender systems in the context of mobile applications and help the new generation of location-based services including Foursquare to engage further with their users.

\section{Acknowledgments}
We acknowledge the support of Microsoft Research and EPSRC through grant GALE (EP/K019392). 
{
\small
\bibliographystyle{aaai}
\bibliography{Bibliography,.}
}

\end{document}